\documentclass[sigconf]{acmart}
\usepackage{comment}
\usepackage{multirow}

\AtBeginDocument{%
  \providecommand\BibTeX{{%
    \normalfont B\kern-0.5em{\scshape i\kern-0.25em b}\kern-0.8em\TeX}}}

\copyrightyear{2021}
\acmYear{2021}
\setcopyright{acmcopyright}

\acmConference[MM '21]{Proceedings of the 29th ACM International Conference on Multimedia}{October 20--24, 2021}{Virtual Event, China}
\acmBooktitle{Proceedings of the 29th ACM International Conference on Multimedia (MM '21), October 20--24, 2021, Virtual Event, China}
\acmPrice{15.00}
\acmDOI{10.1145/3474085.3475281}
\acmISBN{978-1-4503-8651-7/21/10}

\acmSubmissionID{610}

\begin{document}
\fancyhead{}
    \title{CONQUER: Contextual Query-aware Ranking for \\ Video Corpus Moment Retrieval}

\author{Zhijian Hou}
\affiliation{%
  \institution{City University of Hong Kong}
}
\email{zjhou3-c@my.cityu.edu.hk}

\author{Chong-Wah Ngo}
\affiliation{%
  \institution{Singapore Management University}
}
\email{cwngo@smu.edu.sg}

\author{W. K. Chan}
\affiliation{%
  \institution{City University of Hong Kong}
}
\email{wkchan@cityu.edu.hk}


\begin{abstract}
  This paper tackles a recently proposed Video Corpus Moment Retrieval task. This task is essential because advanced video retrieval applications should enable users to retrieve a precise moment from a large video corpus. We propose a novel CONtextual QUery-awarE Ranking~(CONQUER) model for effective moment localization and ranking. CONQUER explores query context for multi-modal fusion and representation learning in two different steps. The first step derives fusion weights for the adaptive combination of multi-modal video content. The second step performs bi-directional attention to tightly couple video and query as a single joint representation for moment localization. As query context is fully engaged in video representation learning, from feature fusion to transformation, the resulting feature is user-centered and has a larger capacity in capturing multi-modal signals specific to query. We conduct studies on two datasets, TVR for closed-world TV episodes and DiDeMo for open-world user-generated videos, to investigate the potential advantages of fusing video and query online as a joint representation for moment retrieval.
\end{abstract}

\ccsdesc[500]{Information systems~Multimedia and multimodal retrieval}
\ccsdesc[500]{Information systems~Video search}

\keywords{\small{Moment Localization With Natural Language; Cross-Modal Retrieval}}


\maketitle


\section{Introduction}

Impressive progress has been witnessed for Video Retrieval~(VR)~\cite{gabeur2020multi,chen2020fine,dong2019dual,awad2016trecvid} or Single Video Moment Retrieval~(SVMR)~\cite{cao2020strong,qu2020fine} tasks, where a natural language query is matched with either a video from a large corpus or a moment within a video. However, those two tasks are not entirely realistic. Current VR benchmarks~\cite{xu2016msr,rohrbach2015dataset,berns2019v3c1} trim video clips by human intervention from the original lengthy video to form the video corpus. Current SVMR benchmarks~\cite{anne2017localizing,gao2017tall,krishna2017dense} assume a priori ground-truth video and consider only visual source. This paper tackles a more realistic \textit{Video Corpus Moment Retrieval}~(VCMR) task. For a natural language query, the system aims to retrieve the ground-truth video from the whole corpus and predict the moment with high Intersection-of-Union~(IoU) with the ground-truth moment. It is a challenging problem because a video corpus typically contains many videos, each video contains many possible moments, queries are open-ended, and moments are query-dependent and vary in time duration.

\begin{figure}
\footnotesize
\begin{center}
\includegraphics[width=0.95\linewidth]{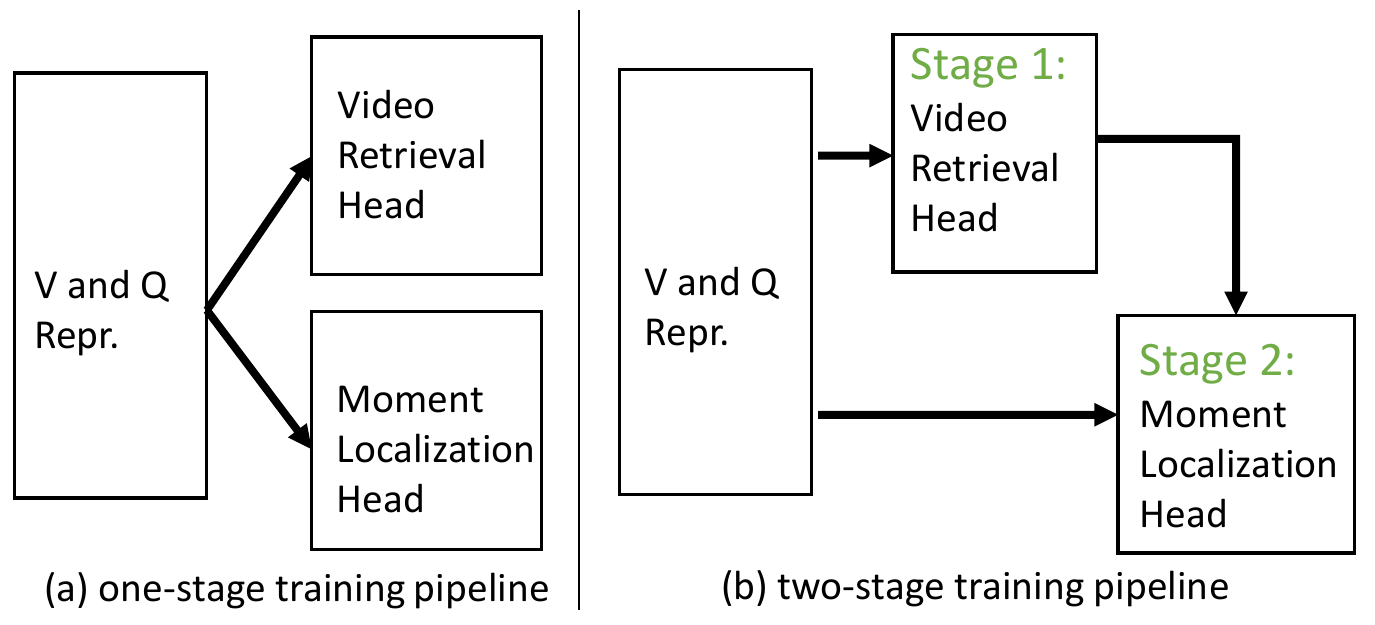}   
\end{center}
\caption{Comparisons between one-stage and two-stage training pipeline. V and Q Repr. means the video and query representations extracted from the backbone. One-stage training pipeline uses a model to train a video retrieval head and a moment localization head simultaneously whereas two-stage training pipeline uses the first model to train a video retrieval head and the second model to train a moment localization head.}
\label{fig:intro_comp}
\end{figure}

State-of-the-art models~\cite{lei2020tvr,li2020hero} adopt the one-stage training pipeline with parallel heads as shown in Fig.~\ref{fig:intro_comp}a. Once the backbone produces the video and query representations, those representations are inputted to its video retrieval head to output VR scores and its moment localization head to output the start and end probability distributions for each clip index. Nonetheless, one-stage training pipeline has inherent limitation on precise moment localization. Current moment localization head integrates each clip and query feature by late-fusion to generate 1D query-clip similarity scores and then uses a 1D Convolution to predict the start and end score distributions. The similarity scores are merely summarized through cosine similarity, and the convolution operator only captures local temporal dependencies within a sliding window. Early fusion, which alters the video representation based on query attention for similarity scoring, is known to be more effective~\cite{liu2018attentive,liu2018cross}. Nevertheless, as video representation can only be generated when query is known, videos cannot be indexed for real-time search. Furthermore, early fusion usually consumes considerable inference time and memory space, causing it less preferable.

In general, early fusion of video and query feature involves two key steps. First, the multi-modality features in a video are fused depending on query context. In the literature, multi-modal fusion is practically carried out by weighted sum of different modality features. Before the deep-learning era, there have been numerous empirical studies conducted to investigate query-dependent fusion~\cite{DBLP:journals/tcsv/PengN06,li2013query,wilkins2010properties}. These works include classification of user information need into visual or text oriented queries and derivation of optimal fusion weights for different query classes~\cite{yan2004learning}. These efforts, however, are in no avail as practically average fusion is known to perform better in some large datasets despite the inherent limitation. The second step is the coupling of video and query features as a joint feature, which we refer to query-aware feature in this paper, for similarity scoring. Since the deep-learning era, especially due to the popularity of attention mechanism~\cite{luong-etal-2015-effective} and transformer~\cite{VaswaniSPUJGKP17}, learning query-aware or question-aware features have been explored, for example, in multimedia question-answering~\cite{antol2015vqa,lei-etal-2018-tvqa}.

\begin{figure*}
\footnotesize
\begin{center}
\includegraphics[width=\linewidth]{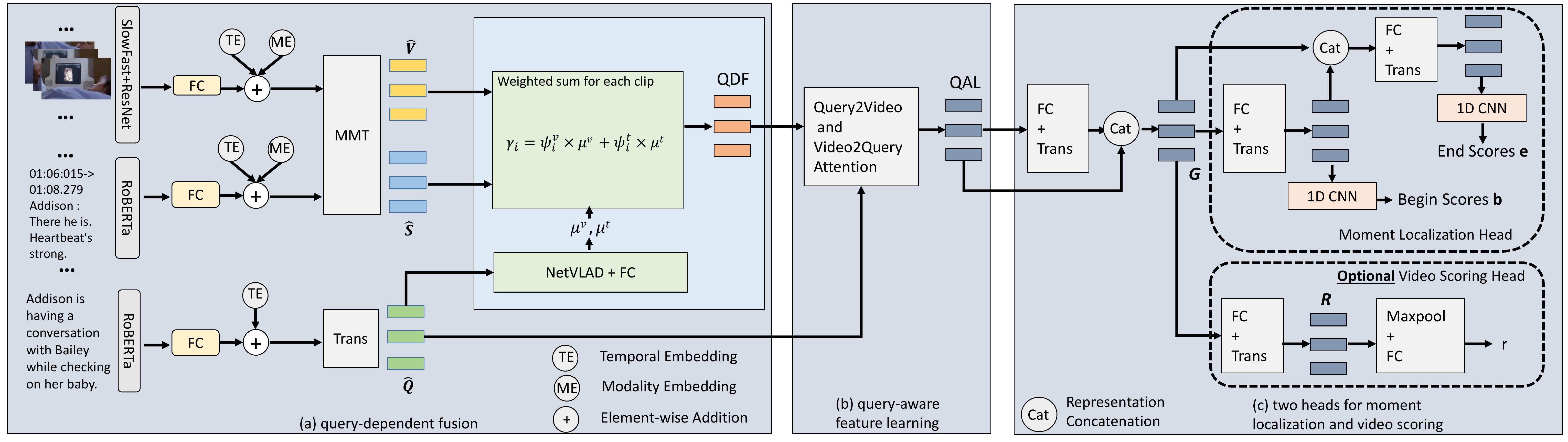}
\end{center}
\caption{CONQUER: (a) query-dependent fusion, (b) query-aware feature learning, (c) two heads for moment localization and video scoring. \textit{Trans = Transformer}. For illustration, each video (query) is assumed to have three clips (tokens). }
\label{fig:stage_two}
\end{figure*}

To the best of our knowledge, both steps of early fusion, i.e., query-dependent fusion (QDF) and query-aware feature learning (QAL), have never been fully explored and jointly exploited. For example, in video search, QDF is a common practice for multi-modal fusion but QAL is seldom employed due to requirement of long inference time. This paper addresses the problem, specifically the integrated use of QDF and QAL, in the context of VCMR. The task is more complex than VR because the query context is expected to couple locally with video for moment localization. We present a new architecture, composed of various off-the-shelf sub-networks~\cite{gabeur2020multi,arandjelovic2016netvlad,SeoKFH17,VaswaniSPUJGKP17,lei2020tvr}, to study the problem. The architecture, as depicted in Fig.~\ref{fig:stage_two}, connects QDF and QAL in two separate steps. QDF derives fusion weights from user query for derivation of video feature. The weights represent the perspective of a query on how different modalities of video content should be composed as a feature for further processing. QAL, subsequently, couples the QDF derived video feature with user query into a single feature for retrieval. We argue that this QDF+QAL derived representation is user-centered and has more potential to answer diverse types of queries. In addition, the representation is more resilient for encapsulating different levels of details in a progressive manner. Specifically, this refers to the processing pipeline where a feature locally extracted from a clip is continuously evolved when being gradually embedded with video and query contexts at different steps from QDF, QAL to a representation readily for moment localization. 

The presented work is envisioned to be a two-stage retrieval pipeline as in Fig.~\ref{fig:intro_comp}b, where the 1st-stage VR search engine supplies top-$k$ video candidates for the proposed architecture to perform moment localization and ranking. The main contribution of this paper is the proposal of a novel learning-based architecture, CONQUER, and the provision of empirical insight on joint exploitation of QDF and QAL in two steps. We study the performance of CONQUER on both the closed-world (i.e., TVR~\cite{lei2020tvr}) and open-world (i.e., DiDeMo~\cite{anne2017localizing}) datasets to show how retrieval performance is improved when video features are gradually evolved to capture query context.

\section{Related Work}
\noindent\textbf{Video Retrieval.} The task ranks videos among corpus based on the similarity with a language query. Standard pipeline learns global embedding of video and query feature separately, maps those two features to joint embedding space, and finally computes similarity scores by cosine distance. Some recent works tackle only visual context and propose separate
embedding for each part-of-speech~\cite{wray2019fine}, or multi-level
encoding~\cite{dong2019dual} and hierarchical graph reasoning~\cite{chen2020fine} for both visual and text modalities, others focus on multi-modal context and propose mixture of embedding experts~\cite{miech2018learning}, collaborative experts~\cite{LiuANZ19}, multi-modal transformer~\cite{gabeur2020multi}. Unlike those works, we aim to retrieve the moment rather than the whole video.

\noindent\textbf{Singe Video Moment Retrieval.} The task localizes a video segment by a distinct and describable sentence from a video. One kind of dichotomy is late-fusion~\cite{anne2017localizing} and early-fusion~\cite{gao2017tall,liu2018attentive,liu2018cross,ghosh-etal-2019-excl,lu2019debug,yuan2019find,zhang2019exploiting,cao2020strong,qu2020fine}. Late-fusion approach computes offline query-agnostic video feature while early-fusion approach computes query-aware video features. Due to better localization performance, the majority of papers adopt the early-fusion approach. Another kind of dichotomy is proposal-based~\cite{anne2017localizing,gao2017tall,HendricksWSSDR18} and proposal-free~\cite{ghosh-etal-2019-excl,lu2019debug,yuan2019find,zhang2019exploiting,qu2020fine,cao2020strong}. Proposal-based approach first generates handcraft heuristic~\cite{HendricksWSSDR18} or sliding window~\cite{anne2017localizing,gao2017tall} proposals while proposal-free approach directly predicts start and end timestamps. Recent papers mainly follow the proposal-free approach due to less computation cost and better results. Unlike those works, we aim to perform retrieval from a large corpus rather than a single video.

\noindent\textbf{Video Corpus Moment Retrieval.}  Compared to VR and SVMR, relatively few researches -- XML~\cite{lei2020tvr}, HERO~\cite{li2020hero}, HAMMER~\cite{zhang2020hierarchical} -- are conducted for this problem. These works mainly explore cross-modal attention~\cite{luong-etal-2015-effective} or transformer~\cite{VaswaniSPUJGKP17} to fuse multi-modal features and provide video context for derivation of frame or clip-level features. As most of these works~\cite{lei2020tvr,li2020hero} target for one-stage retrieval (as in Fig. 1a), query context is only lightly used with little computation overhead to provide basic information for moment localization. For example, XML~\cite{lei2020tvr} applies self-attention to learn two query vectors that respectively capture the visual and textual attentions of user query words. During retrieval, the two online query vectors evaluate text and visual query-clip similarities separately to produce probability scores indicating moment locations. We argue that the late fusion of video and query features for computing similarity as in XML~\cite{lei2020tvr} does not fully exploit query context because video features are derived independent of query. Similarly, in HERO~\cite{li2020hero}, query context is not exploited for multi-modal fusion but, instead, is embedded in a joint space as the video feature for video and clip similarity measures. Different from XML and HERO which employ ConvSE~\cite{lei2020tvr} for moment localization, HAMMER treats VCMR as a frame classification problem and tags each frame in a video with labels to facilitate the prediction of moment span. Like CONQUER, HAMMER is envisioned as a two-stage retrieval pipeline with the 2nd-stage performs only for top-ranked videos. Nevertheless, HAMMER also does not exploit the query context for early fusion with video features. In this paper, we adopt HERO~\cite{li2020hero} as the 1st-stage search engine for CONQUER. HERO is pre-trained on HowTo100M instructional videos ~\cite{miech2019howto100m} and TVR~\cite{lei2020tvr} TV episodes training data, and reported the state-of-the-art performances on datasets such as TVR~\cite{lei2020tvr} and DiDeMo~\cite{anne2017localizing}.

\section{Problem Definition}
\label{subsection:def}
Given a query \textbf{Q}, the problem of VCMR is to locate a moment \textbf{M} in a video \textbf{V} retrieved from a large dataset $D$. The query \textbf{Q} = [$q_1$, $q_2$, ..., $q_m$] is a textual query composed of $m$ tokens. Similarly, a video \textbf{V} = [$c_1$, $c_2$, ....,$c_n$] is composed of $n$ non-overlapping clips and additionally associated with the text descriptions such as subtitles or ASR (automatic speech transcript). Each description is timestamped but does not necessarily align to clip boundaries. An example of a one-second-long subtitle is ``Chase: That's all, is it?" which states the person name and conservation content. Each clip is associated with text descriptions that have time-overlapped with the clip. Hence, a clip $c_i$ is visually and textually described by two different modalities. A moment \textbf{M} is a subsequence of frames in a video \textbf{V}, spanning one or multiple clips. In VCMR, each query is paired with one moment as the only ground-truth answer. Hence, the goal is to rank the moment as top as possible such that the user can quickly browse and locate the right answer of a query. Different from VR which concerns the retrieval of video that contains query moment from $D$, VCMR also requires the moment of that video to be retrieved and with sufficient temporal overlap with the ground-truth.

VCMR is more complex for requiring the localized analysis of clip-level information for moment extraction. The retrieval speed also decreases linearly with the number of clips in a video, |\textbf{V}|, and the number of videos, |$D$|, in a dataset. Hence, a two-stage pipeline is generally helpful in filtering irrelevant videos from clip-level analysis. To this end, given the top-$k$ videos where $k \ll |D|$, the problem of VCMR is to extract the potential moments from $k$ videos and rank them based on their similarity to \textbf{Q}.

\section{CONQUER}
\label{section:conquer}
CONQUER entangles video clips and query progressively to encompass different level of details in two subsequent steps. As depicted in Fig.~\ref{fig:stage_two}, the two steps are query-dependent fusion (QDF) to combine within-clip modalities and query-aware learning (QAL) to learn new representation that aggregates clip and query information. The query-aware feature is utilized by two heads for moment localization and video scoring. Inter-clip and inter-token contextualization is also performed along with each step and head. Table~\ref{table:notation} summarizes the notations used by CONQUER at different steps to represent the video and query features. For simplicity, the bold notation indicates video or query level features, while the notation with subscript refers to clip or token level features. The various off-the-shelf networks being leveraged to derive the features are also listed in the table~\ref{table:notation}.

\begin{table*}
\footnotesize
\caption{Notations used by CONQUER.}
\label{table:notation}
\begin{center}
\begin{tabular}{l | c | c | l }
\toprule
& Video & Query & Network \\
\midrule
\multirow{2}{6em}{Input feature} & Visual: $\textbf{V}=[ v_1,..., v_n ] \in~ \mathbb{R}^{ n \times 4352}$  &  & SlowFast~\cite{feichtenhofer2019slowfast}, ResNet~\cite{he2016deep} \\
&Textual: $\textbf{S}=[s_{1},...,s_n] \in~ \mathbb{R}^{n \times 768}$ & $\mathbf{Q}=[q_1,....,q_m] \in~ \mathbb{R}^{m \times 768} $  & RoBERTa~\cite{corr/roberta} \\
\hline
\multirow{3}{6em}{Query-dependent} & Visual: $\mathbf{\widehat{V}} =[\psi_{1}^{v},...,\psi_{L_v}^{v}] \in~ \mathbb{R}^{L_{v} \times 768}$  &  & Multi-Modal Transformer~\cite{gabeur2020multi}  \\
&Textual: $\mathbf{\widehat{S}}=[\psi_{1}^{t},...,\psi_{L_v}^{t}]\in~ \mathbb{R}^{L_{v} \times 768}$  & $\mathbf{\widehat{Q}} =  [\phi_{1}, ..., \phi_{L_{q}}] \in~ \mathbb{R}^{ L_{q} \times 768}$ &   Transformer~\cite{VaswaniSPUJGKP17}  \\
& QDF:  $\gamma_i =  \psi^{v}_{i} \times \mu^{v} + \psi^{t}_{i} \times \mu^{t} \in~ \mathbb{R}^{768} $ & & NetVLAD~\cite{arandjelovic2016netvlad}  \\
\hline
\multirow{4}{6em}{Query-aware} & QDF: $\gamma_i \in~ \mathbb{R}^{768} $    & \multirow{4}{4em}{---}  & \multirow{4}{16em}{Bi-directional Attention Flow~\cite{SeoKFH17}}   \\
 & V2Q: $\eta_{i} \in~ \mathbb{R}^{768}$    &  &   \\
 & Query contextualized: $\gamma_i \odot \eta_{i} \in~ \mathbb{R}^{768}$    &  &   \\
 & Video contextualized: $\gamma_i \odot q_v \in~ \mathbb{R}^{768} $  &  &   \\

\hline
\multirow{2}{6em}{Moment Localization} & $ \mathbf{b}= [b_1,...,b_{n}] \in~ \mathbb{R}^{n}$ &\multirow{2}{4em}{---} & Transformer~\cite{VaswaniSPUJGKP17}    \\
 & $ \mathbf{e}= [e_1,...,e_{n}] \in~ \mathbb{R}^{{n}}$    &  & ConvSE~\cite{lei2020tvr}   \\
\bottomrule
\end{tabular}
\end{center}
\end{table*}

\subsection{Query-dependent Fusion (QDF)}
\label{subsection:backbone}

Each video clip is described by visual and textual modalities using the deep models stated in Table~\ref{table:notation}. Instead of concatenating both modalities as a long feature vector as the network input~\cite{nie2019multimodal}, we fuse the modalities with linear weighted sum. The weights are adaptively derived dependent on the user query. Specifically, a query such as ``Penny admits Sheldon has said something that she agrees with and haven't thought about before'' emphasizes content of verbal discussion that is expected to be out of video content. In this case, textual features deriving from the textual modality should be given higher priority in fusion. On the other hand, a query such as ``Rachel is shoving a tie into Chandler's mouth'' is more visually than textually appealing, and hence higher visual weight should be assigned. Given $\psi_{i}^{v}$ and $\psi_{i}^{t}$ as the visual and text modalities features of a clip, respectively, the QDF feature is generated by two learnable weights $\mu^{v}$ and $\mu^{t}$, as follows,
\begin{equation}
\label{eq:fused}
 \gamma_i =  \psi^{v}_{i} \times \mu^{v} + \psi^{t}_{i} \times \mu^{t}.
\end{equation}

To enhance the discriminative power of visual and textual modalities, we first convert the input features ${v}_{i}$ and ${s}_{i}$ into the same dimension. Specifically, both modalities are initially undergone linear projection to have the same dimensionality. Following the feature pre-processing of Multi-Modal Transformer~(MMT)~\cite{gabeur2020multi}, each of the projected feature is summed with two vectors that are self-learnt to capture the temporal evolution and modality-specific information. The two vectors named as temporal and modality embedding, respectively, are randomly initialized. As the number of clips in a video varies, a total of $L_v$=100 temporal embeddings are learnt, assuming a maximum of 100 clips per video. Each embedding encodes a time-dependent feature. For a video with less than 100 clips, the non-existing clips are represented by zero vectors summed with the temporal and modality embeddings. Next, to learn the inter-clip relationship, the sets of features are fed in parallel to the MMT transformer to produced the contextualized visual and textual features, $\mathbf{\widehat{V}}$ and $\mathbf{\widehat{S}}$, respectively.  At this phase, the clip-level features not only encode intra-modality signals but are also globally contextualized with video-level information.

Similar transformation is also conducted for the user query, by treating query token as a clip. Note that only temporal embedding is learnt since the query is expressed in text description. A total of $L_q$=30 embedding are learnt, assuming a maximum of 30 tokens in a query. Note that different from the textual modality of a video clip, each query token is encoded as a separate feature and summed with the temporal embedding. The reason is that the content of a user query usually spans across multiple video clips. Keeping each token feature individually instead of collapsing them allows modeling of clip-token relationship, which will be further elaborated in Section~\ref{subsection:neck}. Finally, like video clips, the token-level feature is contextualized by a transformer\footnote{For clarification, Transformer in this work simply refers to the Transformer encoder block in its original paper. } with all other tokens in a query. 

To derive query-dependent weights, the transformed query token features are further aggregated by NetVLAD~\cite{arandjelovic2016netvlad} using a learnt codebook of 32 clusters or code words. The resulting vector globally represents the entire user query Q, and captures the residuals between the query and codebook. Such encoded feature has been proven to be more discriminative as reported in~\cite{arandjelovic2016netvlad}. We add a 1-layer fully-connected network with softmax on top of NetVLAD to learn the query-dependent weights, i.e., $\mu^{v}, \mu^{t} \in~ \mathbb{R}^{1}$~(see Eq.~\ref{eq:fused}). The adaptively fused and globally contextualized multi-modal feature is expected to capture a good mixture of two different modalities prior on the query expectation.

\subsection{Query-aware Feature Learning (QAL)}
\label{subsection:neck}
Up to now, the clip-level fusion reflects the relative importance between two modalities, as adaptively determined based on the query nature. To couple both video and query features for moment localization, we adopt a bi-directional memory-less attention mechanism proposed in~\cite{SeoKFH17}. The mechanism appends the QDF feature~($\gamma_i$) of each video clip in the context with a set of query-aware features. Specially, the appended features are the result of performing attention in two directions: element-wise weighting clip features by query tokens (i.e., query-to-video attention or Q2V) and vice versa (i.e., video-to-query attention or V2Q). The core of attention is a similarity matrix $\in~ \mathbb{R}^{L_{v} \times L_{q} }$, in which an element $a_{i,j}$ represents the similarity between $i$-th clip and $j$-th token, learnt by,
\[ a_{i,j} = \mathbf{w_1} \cdot \gamma_i + \mathbf{w_2} \cdot \phi_{j} + \mathbf{w_3} \cdot (\gamma_i \odot \phi_{j} ),\]
where $\mathbf{w_1, w_2, w_3} \in \mathbb{R}^{768} $ are trainable vectors and $\odot$ represents element-wise multiplication. V2Q determines the relevance of query tokens to a video clip. The attention weight on $j$-th token by $i$-th clip is denoted as $p_{i,j}$ and each attended query vector $\eta_{i}$ of the clip is a linear weighted sum of query-token features, as follows,
\[ p_{i,j} = \frac{ e^{a_{i,j}}}{\sum_{j=1}^{L_q} e^{a_{i,j}}}, \quad \mathbf{\eta_{i}} =\sum_{j=1}^{L_q}\phi_{j}p_{i,j}. \]
Q2V plays attention to the clips that are more similar to one of the query tokens. In this way, clips that are not similar to any token play a relative minor role in the later phase of feature aggregation. Let $b_i$ as the maximum similarity score among all the tokens to $i$-th clip. The attended video vector $q_v$, which is a weighted sum of clip features with respect to query context, is calculated as the follows,
\[ b_{i} = \max_{1 \leq j \leq L_q}a_{i,j}, \quad  p_{i} = \frac{ e^{b_{i}}}{\sum_{j=1}^{L_v} e^{b_{i}}}, \quad q_v = \sum_{i=1}^{L_v}\gamma_{i}p_{i}.\]
The query-aware features can be appended to each query-dependent clip feature, for example, by concatenating $\gamma_i$ with $\eta_{i}$ from V2Q and $q_v$ from Q2V. Nevertheless, as $q_v$ is a vector global to all clips, we follow the suggestion in~\cite{SeoKFH17} to append the following triplets ($\eta_{i}$, $\gamma_i\odot \mathbf{\eta_{i}}$, $\gamma_i\odot q_v$) as QAL feature for each clip.

\subsection{Moment Localization (ML) Head}
The ML head essentially extracts a sequence of clips $[c_b,..,c_e]$, where $b \leq e$, as the query moment. This is conducted by estimating the begin and end score distributions of a moment, respectively, for each clip based on its QAL feature. As a pre-processing, we first contextualize QAL features with a transfomer to learn the inter-clip relationship. The transformed features are concatenated with the features in the previous layer as $\mathbf{\emph{G}} \in \mathbb{R}^{L_v \times 3072}$ to reduce the information loss caused by contextualization. 

To estimate the ``begin'' probabilities of clips, a second round of transformation is performed before feeding the feature to ConvSE~\cite{lei2020tvr}, which is a 1D CNN that produces a score distribution \textbf{b}. To couple the estimation of begin and end probabilities, a third round of transformation is further performed before feeding to ConvSE for prediction of ``end'' score distribution \textbf{e}. For details, please refer to Fig.~\ref{fig:stage_two}(c) for the exact workflow of ML head.

\subsection{Video Scoring (VS) Head}
CONQUER includes an optional head for scoring the similarity of a video to the user query. The head is optional since the similarity can also be derived by consolidating \textbf{b} and \textbf{e} score distributions. Like ML head, the video scoring (VS) head also contextualizes QAL feature with two transformers. The resulting clip features $\mathbf{\emph{R}} \in \mathbb{R}^{L_v \times 768}$ are max-pooled and then undergone linear regression to predict video similarity score.

\subsection{Training Strategy: Shared Normalization}
CONQUER is trained end-to-end to optimize the network parameters, by minimizing the moment cross-entropy loss for the ML head and optional video cross-entropy loss for the VS head. Instead of minimizing the moment loss for each video individually, the softmax operator is shared between the positive video and selected negative videos. In other words, the \textbf{b} and \textbf{e} from different videos are considered jointly such that the learnt begin and end scores are comparable across multiple videos. The technique is referred to as shared normalization originally proposed in~\cite{GardnerC18} for document-level reading comprehension. The begin and end timestamps of a ground-truth moment are first floored and ceiled, respectively, to align with the index of the clip. Denote the aligned ground-truth moment as $\mathbf{M_{gt}}$ = [$c_{b_{gt}},..,c_{e_{gt}}$], where $b_{gt}$ and $e_{gt}$ are the aligned begin and end clip index inside the ground-truth video. The shared softmax normalization produces a begin probability distribution for multiple videos in a mini-batch, as follows,
\begin{equation}
\label{eq:moment-cross-entropy}
   P_{begin} = softmax([\mathbf{b}^{pos},\mathbf{b}^{neg_1}; \mathbf{b}^{neg_2}; \ldots]), 
\end{equation}
where the superscript indicates whether the video in a mini-batch is positive or negative example. The loss of begin timestamp is $-\log P_{begin}{(b_{gt})}$~(see Eq.~\ref{eq:moment-cross-entropy}). The shared normalization is also computed similarly for $P_{end}$ as the end probability distribution. The moment cross-entropy loss is the sum of the begin and end timestamp losses.
The loss function for VS head is also computed in a similar way using shared normalization, which involves multiple videos in a mini-batch. The video cross-entropy loss is the negative log-likelihood of the positive video similarity score.

Note that the negative examples are not sampled in random but according to the video rank list provided by the 1st-stage search engine. To increase the sample size while emphasizing more on hard negative examples, empirically, we sample videos up to a search depth of $d=p+500$, where $p$ is the rank of the ground-truth video. Please refer to Appendix A for further details.

\subsection{Discussion}
Although CONQUER is built on top of various networks in the literature, the integration novelly explores a rigorous approach of coupling query context in video representation learning. Compared to XML~\cite{lei2020tvr} which uses attention model and HERO~\cite{li2020hero} as well as HAMMER~\cite{zhang2020hierarchical} which uses cross-modal transformer as encoders, CONQUER which learns temporal and modality embeddings simultaneously as in MMT~\cite{gabeur2020multi} can model temporal and modality signals more effectively. The use of NetVLAD~\cite{arandjelovic2016netvlad} also makes CONQUER different from XML. Specifically, VLAD and NetVLAD are known to be a powerful feature representation technique for large-scale retrieval~\cite{DBLP:conf/cvpr/JegouDSP10,arandjelovic2016netvlad}. In our work, NetVLAD is utilized to enrich query representation with contextually relevant word clusters before learning query-dependent weights. This is also largely different from XML which performs self-attention to learn visual and textual dependent queries for multi-modality similarity scoring. In addition, while ConvSE~\cite{lei2020tvr} is also used in XML and HERO, the input to ConvSE is a sequence of similarity scores. In contrast, CONQUER input the learnt QDF+QAL features contextualized with transformers as input directly to ConvSE. As the inputs are more discriminative than similarity scores, better performance from ConvSE is expected. Finally, the use of shared normalization~\cite{GardnerC18} is also unique in CONQUER, where the hard negative videos can be sampled from top-$d$ videos for model training. Different from XML and HERO which only uses ground-truth videos for training, this objective function allows CONQUER to distinguish the moment timestamp within the ground-truth video and all the timestamps in hard negative videos.

\section{Moment Retrieval}
CONQUER parses each of the top-$k$ videos recommended by the 1st-stage search engine, and outputs \textbf{b} and \textbf{e} score distributions for each video. By outer product of normalized score distributions, $\mathbf{\hat{b}}$ and $\mathbf{\hat{e}}$, within each video, a matrix is formed where each element indicates the probability of a moment candidate. We post-process the matrix to eliminate short and lengthy moment candidates from consideration. Non-maximum suppression (NMS)~\cite{lin2018bsn} is then performed to further remove candidates with lower probability while having more than 70\% of IoU with an existing moment.

The surviving moment candidates of different videos are subsequently ranked and presented to the user. There are, nevertheless, three possible ways of quantifying moment similarity for ranking. The most \emph{general} way is by multiplying the moment probability obtained by CONQUER with the video similarity score inherited from the 1st-stage search engine. This way is general for naturally leveraging the scores from the two-stage pipeline for retrieval. A more \emph{exclusive} way is by forgetting the video score from the 1st-stage search engine, and directly multiplying the moment probability with the video similarity produced by the VS head of CONQUER. Instead of prior on a global similarity score, a \emph{disjoint} way is to simply rank moment candidates by their probabilities. We experiment all three different ways. In short, denote $r^1$ and $r^2$ as the video similarity score obtained from 1st and 2nd stage retrieval respectively for the $i$-th moment \textit{\textbf{M}}, the three ways of similarity measure are summarized as
\begin{equation}
\label{eq:similarity}
\begin{aligned}
\mbox{General:}  & \quad   \hat{b_i} \times \hat{e_i} \times r^1, \\
\mbox{Exclusive:} & \quad   \hat{b_i} \times \hat{e_i} \times r^2, \\
\mbox{Disjoint:}  &  \quad  b_i + e_i, \\
\end{aligned} 
\end{equation}
where $\hat{b_i}$ and $\hat{e_i}$ are normalized begin and end scores within the video at a given begin and end index for the $i$-th moment \textit{\textbf{M}}, respectively. Note that \emph{disjoint} does not use the normalized probabilities because the normalized scores are not comparable when not being prior on the video similarity score.

\section{Experiment}

\subsection{Dataset}

\begin{table}
\footnotesize
\caption{Dataset statistics showing the number of videos in different splits.}
\label{table:dataset_statistics}
\begin{center}
\begin{tabular}{l | l l l  l }
\toprule
Dataset & Training & Validation & Testing \\
\midrule
TVR & 17,435 & 2,179 & 1,089 \\
\hline
DiDeMo & 8,395 &  1,065 & 1,004 \\
\bottomrule
\end{tabular}
\end{center}
\end{table}

We conduct experiment on two datasets: TVR~\cite{lei2020tvr} and DiDeMo~\cite{anne2017localizing}. The statistics of both datasets are listed in Table~\ref{table:dataset_statistics}. TVR is a closed-world dataset where the videos are drawn from 6 long-running TV shows. Videos are accompanied with timestamped subtitles and are on average 76.2 seconds. There are 5 queries designed for each videos, resulting in 5,445 queries in the testing set. The moment length ranges from 0.29 seconds to 123.02 seconds and on average lasts for 9.1 seconds. In contrast, the DiDeMo videos are open-world, unedited and randomly sampled from YFCC100M Flickr videos~\cite{thomee2016yfcc100m}. The average length is shorter with 54 seconds. Note that the videos in DiDeMo are not accompanied with text descriptions. Instead, ASR is provided by~\cite{li2020hero} for about 50\% of the videos. There are on average 4 queries for each video. The average length for query and moment is 8 tokens and 6.5 seconds, respectively.

While the focus of this paper is on VCMR, we also perform evaluation on VR and SVMR. The evaluation metric follows~\cite{lei2020tvr}, which is the average recall of queries at the search depth of $k$ or R@$\mathit{k}$. For VCMR and SVMR, an additional IoU threshold (0.5 or 0.7) with ground-truth moment is set. Only moment candidates with IoU equal to or more than the threshold are counted as positives. Note that IoU is measured based on the overlapped timestamps between the retrieved and ground-truth moments.

\subsection{Implementation}

Following XML~\cite{lei2020tvr}, we uniformly divide each video into a sequence of 1.5-second clips. As the clip boundaries do not align with timestamps of text descriptions, we simply assign descriptions to clips that have timestamp overlap between them. The assignment is not one-to-one, or precisely, a textual description may associate with consecutive video clips. In moment localization, we adopt the same assumption as~\cite{lei2020tvr} where the prior knowledge of moment length is known to a system. Specifically, the moment length is constrained within [$L_{min}$, $L_{max}$]. The values correspond to the minimum and maximum number of clips in a moment, respectively. Therefore, moment candidates which are shorter or longer than this range are directly pruned from further processing. We empirically set $L_{min}= 1$  and $L_{max}=24$ for TVR dataset, and $L_{min}=3$ and $L_{max}=7$ for DiDeMo dataset in the following experiments. Please refer to Appendix B for further details.

The learning rate is set as 1e-4. There are two cross-entropy loss functions, respectively, correspond to the ML and VS heads of CONQUER. The weights for moment and video loss are set to 1e-2, 5e-2, respectively. Note that we only use the moment loss for general and disjoint similarity scoring function~(see Eq.~\ref{eq:similarity}), whereas we use the combination of the two for exclusive similarity scoring function. We adopt early stopping approach, and compute the sum of R@1 values at IoU=\{0.5, 0.7\} after each training epoch. The training only involves queries whose ground-truth videos can be retrieved within the search depth of 100 by the 1st-stage search engine. This is simply because some queries are ambiguous and could have multiple moments as ground-truths although only one of them is marked as positive. Involving these queries can negatively impact learning effectiveness. There are about 0.5\% of queries being excluded from training. The training is ended when the value of sum consecutively drops for 3 epochs in the validation set. The training time depends on the number of negative videos being sampled. When keeping the number to 3, CONQUER converges in about 7.5 to 9 hours depending on the similarity scoring function for the TVR dataset. The experiments are conducted on one 2080 Ti GPU using AdamX optimizer~\cite{LoshchilovH19}.

\subsection{Ablation Study}

We conduct experiments to justify the contribution of three major components (QDF, QAL and shared normalization) in CONQUER, the setting of top-$k$ and the similarity measure for moment ranking. The experiments are conducted on the validation set of TVR because the ground-truth of the test set is not publicly available. The results are presented in tables~\ref{table:ablation_arch},~\ref{table:function_comparison} and ~\ref{table:inference_number_k}, respectively. As a reference, when considering the top-10 videos returned by HERO, the oracle VCMR performance in terms of recall is 63.07. This is based on the assumption that, for all the queries, the oracle model manages to extract the ground-truth moments from top-10 videos and rank them within the search depth. This result also indicates that, among the 10,895 validation queries, HERO manages to rank the ground-truth videos of 6,872 queries within the search depth of $k=10$. Note that when employing HERO as the 1-st stage search engine, we only use the search result produced by its video retrieval head.

Table~\ref{table:ablation_arch} details the contribution of each component, by ranking the moment candidates extracted from top-10 videos with {\em general} scoring (see Eq.~\ref{eq:similarity}). The baseline is an ``empty'' CONQUER where modalities are transformed by MMT, averagely fused, multiplied by aggregated query features with self-attention~\cite{lei2020tvr}, and finally forwarded to ConvSE for moment localization. When comparing this baseline to CONQUER with either QDF or QAL component, it is noted that QAL plays a more significant role in boosting the performances of both VCMR and SVMR. QDF alone manages to push more ground-truth moments within top-5 depth. The result shows that even a relatively simple way of aggregating video and query features, such as the self-attention adopted by baseline, actually works fairly well in ranking. Compared to baseline, CONQUER, which adopts transformer for bi-directional attention modeling, shows much better performance. When combining both QDF and QAL, substantial improvement is noted when comparing to baseline. Shared normalization also contributes to the performance, showing the strategy of using softmax over a positive video and multiple hard negative videos is practically effective in model training. Overall, there are 2,488 out of 10,895 queries where the moments are correctly extracted with at least 70\% IoU overlap with the ground-truth and ranked at top-1 position for SVMR. Although some of these moments are not ranked within top-5 for VCMR task, the result still indicates the effectiveness of CONQUER, despite still having a big performance gap compared to the oracle result.

\begin{table}
\footnotesize
\caption{CONQUER: Incremental contributions of every step on TVR validation set.}
\label{table:ablation_arch}
\begin{center}
\begin{tabular}{ c c c | c c c l c c c }
\toprule
& &  \multirow{2}{5em}{Shared normalization} &  \multicolumn{3}{c}{VCMR IoU=0.7} & & \multicolumn{2}{c}{SVMR IoU=0.7}  \\
\cline{4-6} \cline{8-9} 
QDF & QAL &  &  R1 & R5 & R10 & & R1  & R5  \\
\midrule
 &  & &5.89 & 13.62 & 18.55  & & 16.92
  & 37.82    \\
\hline
\checkmark &  & &5.88 & 13.78 & 18.8 & & 16.69 & 38.12 \\
\hline
 &  \checkmark &  & 6.98 & 16.24 & 21.04
  & & 21.18  & 42.96 \\
\hline
\checkmark &  \checkmark &  & 7.21 & 16.1 & 21.09
  & & 22.15  & 43.06 \\
\hline
\checkmark &  \checkmark & \checkmark &\textbf{7.76} & \textbf{16.46} & \textbf{21.47}
 & & \textbf{22.84} & \textbf{43.64}  \\
\bottomrule
\end{tabular}
\end{center}
\end{table}

\begin{table}
\footnotesize
\caption{R@1 performance of CONQUER: Impact of similarity measure for moment ranking on TVR validation set.}
\label{table:function_comparison}
\begin{center}
\begin{tabular}{ l | c l  c l c }
\toprule
     & VCMR IoU=0.7 &  & VR  &  & SVMR IoU=0.7 \\
\midrule
General & \textbf{7.76} &  &  29.01 & & \textbf{22.84}  \\
\hline
Disjoint & 7.18  &  & 26.67 & & 21.84 \\
\hline
Exclusive  & 7.02  &  & \textbf{29.29} & & 20.10  \\
\bottomrule
\end{tabular}
\end{center}
\end{table}

Table~\ref{table:function_comparison} contrasts the impact of three similarity measures for ranking of moment candidates. Note that, for {\em general} and {\em disjoint}, CONQUER is trained with ML head only. Comparing {\em disjoint} and {\em exclusive}, the former shows better R@1 for VCMR while the latter shows better R@1 for VR. The result indicates that joint learning of ML and VS heads could hurt the performance of moment localization, despite improving the performance of video search. In SVMR, the fact that {\em disjoint} outperforms {\em exclusive} when ranking moment candidates within the ground-truth video further confirms this finding. The result gives clue to the difficulty of training two ML and VS heads simultaneously, and is consistent with our speculation that one-stage training with two parallel heads could result in suboptimal moment localization. Overall, {\em general}, which computes similarity score prior on the result of 1st-stage search engine, offers the overall best performance in both VCMR and SVMR.

\begin{table}
\footnotesize
\caption{CONQUER: Impact of video pool size to VCMR retrieval and speed performances on TVR validation set. Note that the speed only includes model inference time.}
\label{table:inference_number_k}
\begin{center}
\begin{tabular}{ l | l l l l | l }
\toprule
 & \multicolumn{3}{c}{VCMR IoU=0.7} & &  \\
\cline{2-5}  \
top-k & R1 & R10 & R100 & & speed~(ms) \\
 \midrule
1 &7.24 &18.25 &22.26 & & 3\\
\hline 
5 &7.73 &22.29 & 32.74 & & 8.8\\
\hline 
10 &7.76 &22.49 & 35.17 & & 16.2\\
\hline 
100 &7.77 &22.73 & 37.67 & & 141.9\\
\hline 
200 &7.77 &22.73 & 37.67 & & 283.4 \\
\bottomrule
\end{tabular}
\end{center}
\end{table}

Next, we study the impact of video pool size, i.e., top-$k$, towards VCMR performance. On TVR validation set, the performance indeed saturates quickly with the increase value of $k$ and becomes stable at around $k=10$. From top-1 to top-5, the video retrieval (VR) performance of HERO is improved from 29.01\% to 52.82\% in terms of recall. The VCMR performances of CONQUER at R@1 and R@10 are also boosted accordingly but the performance margin is relatively narrow compared to VR. The smaller degree of improvement is not surprising, since the number of moment candidates increases substantially with a larger video pool size, which also inherently raises the challenge of pushing ground-truth moment to better rank. When further enlarging pool size to beyond $k=5$, the margin of improvement gets smaller. It could imply that, for queries where the ground-truth videos are not managed to be ranked within top-5, the corresponding ground-truth moments are also equally hard to be ranked in better positions. Nevertheless, our result shows that the challenge is not just due to moment localization but also the difficulty of ranking the positively extracted moments when video pool size increases. For example, when performing SVMR on the ground-truth videos of queries whose ground-truth moment ranks are outside of 10, CONQUER manages to rank about 25\% of them with their ground-truth moments within top-5 position. Among them, more than 50\% are ranked within top-3. On the other hand, while the performance saturates rapidly, the VCMR performance of CONQUER is not adversely impacted.  When top-$k$ is increased from $k=10$ to $200$, CONQUER still maintains the same level of performance at R@1 and R@10. In terms of speed efficiency, CONQUER only incurs little computation overhead with an average speed of 142 milliseconds per query even when top-100 videos are considered. Overall, HERO+CONQUER can process almost 30 queries per second when $k=10$.  In the remaining experiments, we set $k=10$ and use {\em general} similarity measure as the default setting, unless otherwise stated.

\subsection{Performance Comparison}

\begin{figure*}
\footnotesize
\begin{center}
\includegraphics[width=0.95\linewidth]{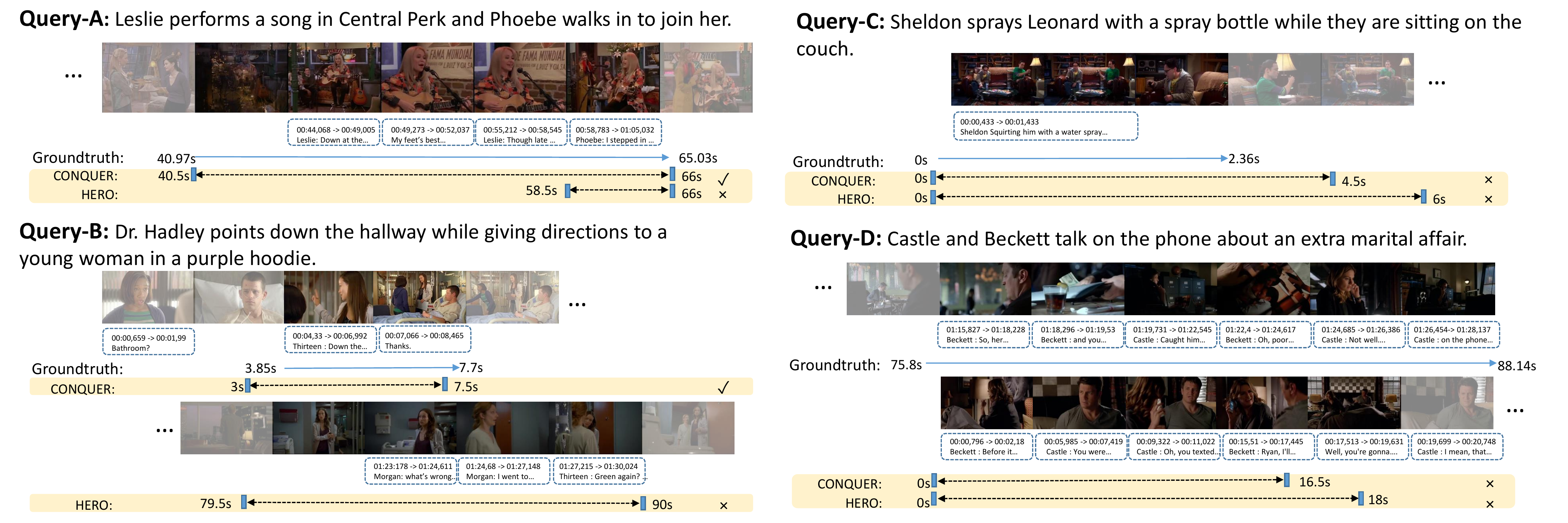}
\end{center}
\caption{Four example queries showing the moments localized by CONQUER and HERO. A tick is given to the correctly localized moment, while a cross indicates negative localization because of IoU $\leq$ 0.7. Best view by zooming to cross-read the query, subtitle and video timeline.}
\label{fig:qualitative}
\end{figure*}

Table~\ref{table:tvr_val_test_comparison} lists the results of various approaches on TVR dataset. CONQUER significantly outperforms all approaches across different recall levels on both validation and testing sets. XML, which contextualizes video-level features with attention models, generally shows lower performance than others which adopt more advanced transformers for contextualizing features. HERO, which is pre-trained on large dataset (Howto100M~\cite{miech2019howto100m}), offers stronger performance than XML and HAMMER. Compared to CONQUER, however, the performance is still limited for not contextualizing video feature with query. The improvement introduced by CONQUER is considered impressive. For example, at R@10, the performance closely approaches the performance of HERO at R@100 on the validation set, and the performance of XML at R@100 on the testing set. 

Figure~\ref{fig:qualitative} shows four typical queries contrasting the performances of CONQUER and HERO. In Query-A, HERO only localizes the short segment where ``Phoebe walks in to join'' while CONQUER also manages to locate the prior segments showing ``Leslie performs a song in Central Perk''. For this query, CONQUER gives higher weight to visual ($\mu^{v}=0.62$). Hence, even Central Perk or singing is not mentioned in the transcript, CONQUER is still able to locate the moment quite precisely. In Query-B, HERO can not capture the query phrase ``a young woman in a purple hoodie'' and locate a moment where Dr. Hadley talks to a woman. In contrast, by intertwining video and query features into a joint feature, CONQUER manages to locate the ground-truth moment where a woman with purple hoodie is clearly visible. In Query-C, both HERO and CONQUER localize a segment that fully cover the target moment. However, both segments also include extra clips that showing ``Sheldon is putting down the spray''. Nevertheless, the moment localized by CONQUER is in higher quality as it includes only the scene where Sheldon holds spray bottle. Query-D shows an example query which is considered challenging because the pharse ``extra marital affair'' is an abstract concept which cannot be easily inferred from subtitle or visual content. For this query, both CONQUER and HERO localize a moment where Castle and Beckett talk about their potential marital affair.

\begin{table}
\footnotesize
\caption{Comparison of VCMR performance (IoU=0.7) with various state-of-the-art approaches on TVR.}
\label{table:tvr_val_test_comparison}
\begin{center}
\begin{tabular}{ l | l l l l l l l}
\toprule
  & \multicolumn{3}{c}{Validation} & &  \multicolumn{3}{c}{Testing} \\
\cline{2-4}  \cline{6-8}
Model &R1 &R10 &R100 & &R1 &R10 &R100 \\
\midrule
\hline
XML~\cite{lei2020tvr}&2.62 &9.05 &22.47 & &3.32 &13.41 &30.52\\
\hline
HAMMER~\cite{zhang2020hierarchical}&5.13 &11.38 &16.71 & &- &- &-\\
\hline
HERO~\cite{li2020hero}&5.13 &16.26 &24.56 & &6.21 &19.34 &36.66\\
\hline
\hline
CONQUER~(general) &\textbf{7.76} &22.49 &\textbf{35.17} & &\textbf{9.24}  &\textbf{28.67} &41.98\\
\hline
CONQUER~(disjoint) &7.18 &\textbf{23.0} &33.94 & &8.5  &26.72 &39.12\\
\hline
CONQUER~(exclusive) &7.02 &22.44 &34.69 & &7.33 &26.81 &\textbf{42.19}\\
\bottomrule
\end{tabular}
\end{center}
\end{table}

\begin{table}
\footnotesize
\caption{Results on DiDeMo Testing set.}
\label{table:test_comparison_dedimo}
\begin{center}
\begin{tabular}{ l | l l l c l l l}
\toprule
  & \multicolumn{3}{c}{VCMR IoU=0.5} & &\multicolumn{3}{c}{VCMR IoU=0.7} \\
\cline{2-4} \cline{6-8}
 Model  &R1 &R5 &R10 & & R1 &R5 & R10\\
\midrule
XML~\cite{lei2020tvr} &2.26 &- &10.42 & &1.59 &- &6.77\\
\hline
HERO~\cite{li2020hero} & \textbf{3.37} & 8.79 &13.26 &&2.76 &7.73 & 11.78\\
\hline
\hline
CONQUER~(general)  & 3.31 & \textbf{9.27} & \textbf{13.99} & & \textbf{2.79} & \textbf{8.04} & \textbf{11.9}\\
\hline
CONQUER~(disjoint)  & 3.09 & 8.49 & 12.86 & & 2.66 & 6.93 & 10.27\\
\hline
CONQUER~(exclusive)  & 3.14 &8.16 &12.05 & &2.61 &7.03 &11.02 \\
\bottomrule
\end{tabular}
\end{center}
\end{table}

Table~\ref{table:test_comparison_dedimo} compares the performances on DiDeMo dataset~\cite{anne2017localizing}. Note that only 50\% of videos are supplied with text descriptions, and almost all the queries are visual-oriented. For these reasons, CONQUER only considers visual modality. In other words, the QDF in CONQUER is turned off for this experiment. Nevertheless, HERO\footnote{The HERO result is based on our reproduced result since there is an error in the original HERO code for evaluating the DiDeMo dataset.} still exploits the extracted ASR information and benefits from it. As shown, CONQUER still significantly outperforms XML but the improvement over HERO is not as significant as in TVR. We speculate that, as the videos in DiDeMo are not as correlated as in TVR, QAL in CONQUER cannot take advantage of closed-world knowledge to calibrate video feature with query context. Nevertheless, the result of CONQUER is still better or comparable to that of HERO, despite not exploiting text modality.



\section{Conclusion and Future work}
We have presented CONQUER as a module designated for two-stage moment retrieval. Compared to one-stage retrieval, CONQUER has more capacity to exploit query context for adaptive multi-modal fusion and representation learning. As demonstrated on TVR dataset, CONQUER is particularly effective and introduces a large margin of improvement compared to the current state-of-the-arts. The improvement due to QAL is most significant, and the best result is attained when combining QDF, QAL and shared normalization. While CONQUER obtains overall encouraging results on TVR and DiDeMo datasets, the result also indicates that query context is relatively easier to exploit in close-world than open-world dataset. In terms of retrieval efficiency, with HERO as the 1-st stage search engine, the overall speed of HERO+CONQUER manages to process about 30 queries per second on a dataset with slightly more than 2000 videos of about 46 hours. 

Furthermore, integration with audio is also possible except the increase in computational cost. Currently, the clip duration is set based on the setting in HERO. We do believe there are other better options, for example to have variable clip length that aligns with the subtitle timestamp. We will leave these issues as future work.

\begin{acks}
We thank the reviewers for their helpful feedback. This research was partially supported by the Singapore Ministry of Education (MOE) Academic Research Fund (AcRF) Tier 1 grant, the GRF of Hong Kong RGC
(project no. 11203517), a CityU MF\_EXT (project no. 9678180), and a
CityU SGS Conference Grant.
\end{acks}

\clearpage
\bibliographystyle{ACM-Reference-Format}
\bibliography{egbib}

\clearpage
\appendix

\section*{Appendix}

\section{Additional Information on Training}
\subsection{The 1-st Stage Search Engine}
In order to make CONQUER independent of HERO~(in case of experiment bias), we indeed use a simplified version of CONQUER as the 1-st stage search engine. Referring to Fig.~\ref{fig:stage_two}(a), we still use the MMT and transformer to compute  $\mathbf{\widehat{V}}$, $\mathbf{\widehat{S}}$ and $\mathbf{\widehat{Q}}$ for QDF. NetVLAD is used to derive the fusion weight $\mu^{v}, \mu^{t}$. Same as XML~\cite{lei2020tvr}, we also learn to aggregate $\mathbf{\widehat{Q}}$ into two vectors as query vectors for visual and textual modalities, respectively. The vectors are directly used to compute the video similarities with $\mathbf{\widehat{V}}$, $\mathbf{\widehat{S}}$. The two similarities scores are further weighted sum using the fusion weights $\mu^{v}, \mu^{t}$. Note that QAL and ML heads are not used in this simplified CONQUER.

\subsection{Negative Video Sampling}
\label{sub:negative_sample}

\begin{figure}[b]
\footnotesize
\begin{center}
\includegraphics[width=0.95\linewidth]{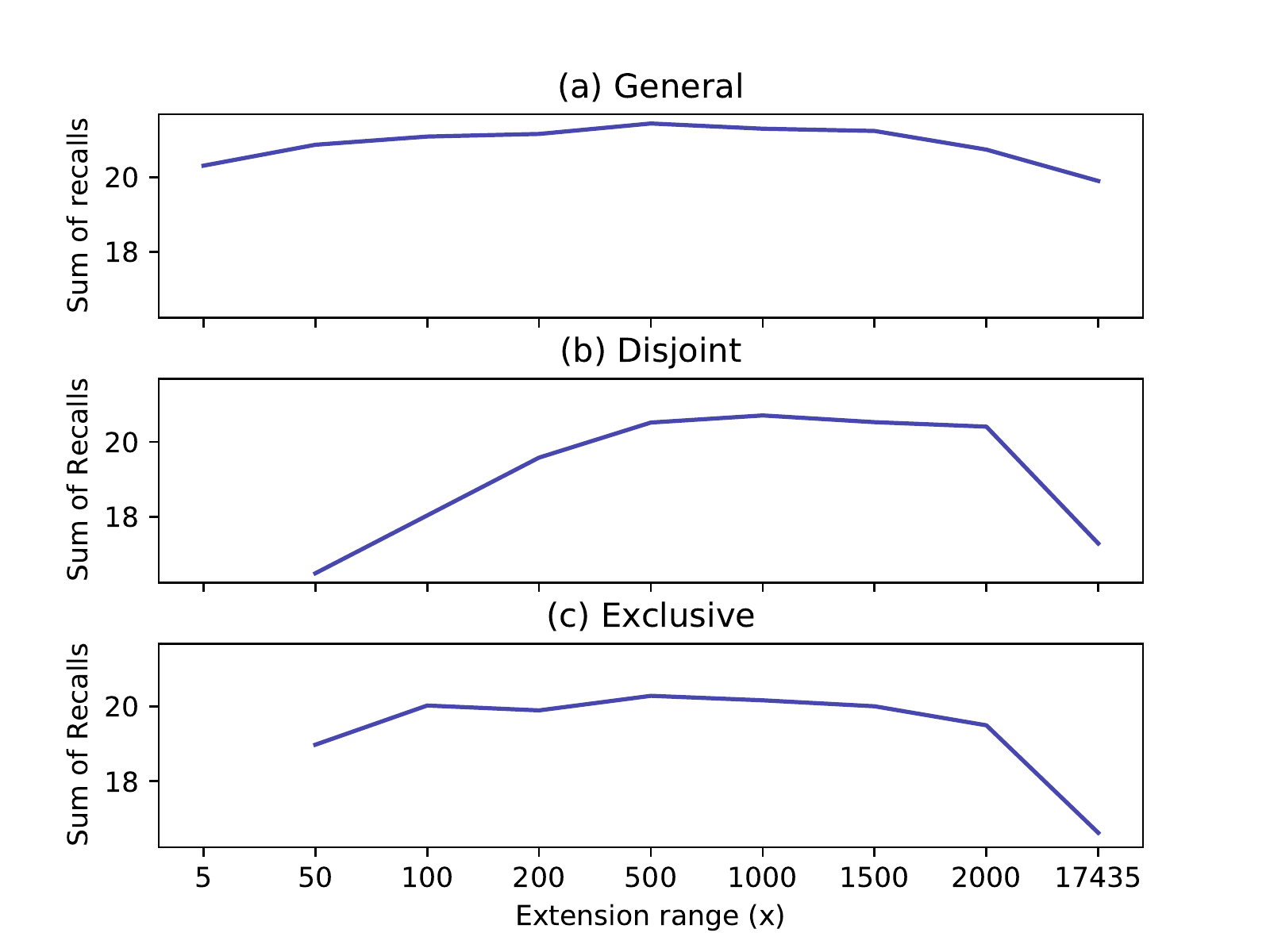}
\end{center}
\caption{Effects of negative video sampling. The y-axis shows the sum of R@1 for both IoU=\{0.5, 0.7\} and x-axis is the extension range of search depth.}
\label{fig:all_ranker}
\end{figure}

During training, we sample videos of a mini training batch up to a search depth of $d=p+x$, where $p$ is the rank of the ground-truth video, and $x$ is the extension range. Fig.~\ref{fig:all_ranker} shows the VCMR result sum curves for different similarity scoring functions when the negative video number is set as 3. As we can see, $x=500$ achieves the optimal performance and is the final adopted setting in this work. Compared with the performance of $x=500$, the performances of randomly sampled negative videos~(see $x=17,435$) are always inferior. It indicates the importance to sample top-ranked videos as negative videos.

\section{The Moment Length Constrains}
\label{sub:moment_length}
Same as XML~\cite{lei2020tvr} and HERO~\cite{li2020hero}, we assume the knowledge of moment length for pruning short and lengthy moments. We use $L_{min}=1$ and $L_{max}=24$, instead of $L_{min}=2$ and $L_{max}=16$ as suggested by HERO for TVR dataset. Table~\ref{table:tvr_length_constrain} provides the experimental results if we set the moment length constrains as HERO, i.e., $L_{min}=2$ and $L_{max}=16$. As shown in Table~\ref{table:tvr_length_constrain}, CONQUER still outperform HERO and XML significantly by using this setting. Note that the setting ($L_{min}=3$ and $L_{max}=7$) on DiDeMo is the same for both HERO and CONQUER.

\section{More Results on TVR}
Table~\ref{table:val_comparison} provides the full set of results comparing CONQUER with XML, HAMMER and HERO. The results include performance for VCMR, VR and SVMR.

\begin{table}[b]
\scriptsize
\caption{Results of CONQUER if using the setting of $L_{min}=2$ and $L_{max}=16$ suggested by HERO for result pruning.}
\label{table:tvr_length_constrain}
\begin{center}
\begin{tabular}{ l | l l l l c l l}
\toprule
     & \multicolumn{4}{c}{VCMR}  &  & \multicolumn{2}{c}{SVMR IoU=0.7} \\
\cline{2-5} \cline{7-8} 
Model  &R1 &R5 & R10 & R100 & &R1 &R5\\
\midrule
XML~\cite{lei2020tvr}  &  2.62 & 6.39 & 9.05 & 22.47 & & 13.89 & 31.11 \\
\hline
HERO ~\cite{li2020hero} &  5.13 & 12.24 &16.26 & 24.56  & & 15.36 & 32.27 \\
\hline
\hline
CONQUER~(general)  &\textbf{7.25} &16.49 & \textbf{21.66} &\textbf{33.88}   & & \textbf{21.87} & 46.18 \\
\hline
CONQUER~(disjoint)  & 6.89 & \textbf{16.88} & 22.38 & 33.29 & & 21.03 & \textbf{46.98} \\
\hline
CONQUER~(exclusive)  & 6.79 &16.35 &21.57 &33.8 & & 19.55 & 43.83  \\
\bottomrule
\end{tabular}
\end{center}
\end{table}

\begin{table}[b]
\tiny
\caption{Full results on TVR validation set.}
\label{table:val_comparison}
\begin{center}
\begin{tabular}{ l | l l l l c l l  c l l}
\toprule
   & \multicolumn{4}{c}{VCMR IoU=0.7} &  & \multicolumn{2}{c}{VR}  &  & \multicolumn{2}{c}{SVMR IoU=0.7} \\
\cline{2-5} \cline{7-8} \cline{10-11} 
 Model &R1 &R5 &R10 &R100 & &R1 &R5 & &R1 &R5\\
\midrule
XML~\cite{lei2020tvr}  &2.62 &6.39 &9.05 &22.47 & &  16.08 & 37.92 & & 13.89 & 31.11 \\
\hline
 HAMMER  &5.13 & - &11.38 &16.71 & & - & - & &  - & - \\
\hline
 HERO ~\cite{li2020hero}  &5.13 &12.24 &16.26 &24.56 &  &   29.01 & 52.82  & & 15.36 & 32.27 \\
\hline
\hline
CONQUER~(general) &\textbf{7.76} &17.22 & 22.49 &\textbf{35.17}  &  &  29.01 & 52.82 & & \textbf{22.84} & 47.72 \\
\hline
CONQUER~(disjoint) &7.18 & \textbf{17.4} & \textbf{23.0} & 33.94  &  &  26.67
 & 55.89 & & 21.84 & \textbf{48.63} \\
\hline
CONQUER~(exclusive) &7.02 &17.03 &22.44 &34.69 &  &  \textbf{29.29}
 & \textbf{56.15} & & 20.1 & 45.12  \\
\bottomrule
\end{tabular}
\end{center}
\end{table}

\end{document}